\documentclass[final, 12pt]{article}
\usepackage[english]{babel}
\usepackage[latin1]{inputenc}
\usepackage{hyperref}
\usepackage{amssymb,amsmath,amsthm}
\usepackage{enumerate}
\usepackage[conditional,light,first,bottomafter]{draftcopy}
\draftcopyName{DRAFT\space\today}{130}
\draftcopySetScale{65}
\usepackage[letterpaper, hmargin=2.3cm,vmargin={2.5cm,3cm}]{geometry}
\usepackage{setspace}
\makeatletter
\renewcommand{\section}{\@startsection%
{section}%
{1}%
{0em}%
{1.7em}%
{1.2em}%
{\normalfont\large\centering\bfseries}}
\renewcommand{\@seccntformat}[1]%
{\csname the#1\endcsname.\hspace{0.5em}}
\makeatother

\numberwithin{equation}{section}
\usepackage{fancyhdr}
\usepackage[usenames]{color}

\numberwithin{equation}{section}
\newtheorem{theorem}{Theorem}[section]

\newtheorem{lemma}{Lemma}[section]

\theoremstyle{definition}
\newtheorem{definition}{Definition}[section]

\newtheorem{condition}{Condition}
\theoremstyle{remark}
\newtheorem{remark}{Remark}[section]

\def\ocirc#1{\ifmmode\setbox0=\hbox{$#1$}\dimen0=\ht0 \advance\dimen0
  by1pt\rlap{\hbox to\wd0{\hss\raise\dimen0
  \hbox{\hskip.2em$\scriptscriptstyle\circ$}\hss}}#1\else {\accent"17 #1}\fi}
\DeclareMathOperator{\re}{Re}
\DeclareMathOperator{\im}{Im}
\DeclareMathOperator{\dom}{dom}

\DeclareMathOperator{\divergence}{div}
\begin{document}
\begin{titlepage}
\title{Correct solvability of hyperbolic Volterra equations with kernels depending on the parameter} 


\author{
\textbf{Romeo Perez Ortiz\footnote{Supported by the Mexican Center for Economic and Social Studies (CEMEES, by its Spanish acronym).}, Victor V. Vlasov\footnote{Supported by the Russian Foundation for Basic Research, project N14-01-00349a and N13-01-00384a}}
\\[6mm]
\small Faculty of Mechanics and Mathematics \\[-1.6mm]
\small Moscow Lomonosov State University \\[-1.6mm]
\small Vorobievi Gori, Moscow, 119991, Russia\\[1mm]
\small\texttt{cemees.romeo@gmail.com}\\[-1mm]
\small\texttt{vlasovvv@mech.math.msu.su}
}
\date{}
\maketitle
\vspace{4mm}
\begin{center}
\begin{minipage}{5in}
  \centerline{{\bf Abstract}} \bigskip  We study the correct solvability of an abstract functional differential equations in Hilbert space, which includes integro-differential equations describing  evolution of thermal phenomena, heat transfer in materials with memory or sound propagation in viscoelastic media. 
 \end{minipage}
\end{center}
\thispagestyle{empty}
\end{titlepage}
\section{Introduction}\label{introduction}
We study functional differential and integro-differential equations with unbounded operator coefficients in a Hilbert space. The main part
of the equation under consideration is an abstract hyperbolic-type equation, disturbed by terms involving Volterra operators. These equations can be regarded as an abstract form of the Gurtin-Pipkin equation (see \cite{{PI}, {PG}} for more details), which describes  evolution of thermal phenomena, heat transfer in materials with memory or sound propagation in viscoelastic media.  It also arises in homogenization problems in porous media (Darcy law). Countless examples about Gurtin-Pipkin type equation  are studied in \cite{GMJ}. It is shown that the initial boundary value problems for these equations are well-solvable in Sobolev spaces on the positive half-axis (see, for instance, \cite{{VR}, {VW}, {KVW}, {VRShamaev}, {RPV}, {vvvlasov}}). 

For a self-adjoint positive operator $A$ considered, we can take, in particular, the operator $Au=-\mu \Delta u-(\lambda+\mu)\triangledown(\divergence u)$ where $\mu$, $\lambda$ are the {\em Lame coefficients} or $A=-\Delta$ with different boundary conditions (for more details,  see \cite{{JME}, {JM}}). Actually, there is an extensive literature on abstract integro-differential equations (see \cite{{PI}, {P},  {PG}, {VR}, {KVW}, {VRShamaev}, {RPV},  {LM}, {vvvlasov}} and the references therein).

Vlasov and Rautian in \cite{VR} established well-defined solvability of initial boundary value problems in weighted Sobolev space on the positive semi-axis for the case $\xi=1$. In the present paper we establish also the well-defined solvability, but for the case $\xi \in (0, 1)$. Unlike the works \cite{{VR}, {KVW}, {VRShamaev}}, we study the correct solvability of hiperbolic Volterra equations with kernels depenging on the parameter. 
In our case, the extreme cases $\xi=1, \xi=0$ are included. The present work is a natural extension of the results \cite{{VR},  {KVW}, {VRShamaev}}.  

It is important to mention here that in the paper \cite{RPV} we provided a theorem about the well-defined solvability for the case $\xi \in (0, 1)$ and there we mentionated that the result was the same as in \cite[Theorem 1]{VR}. In reality the result is more general and different. In the section \ref{proof-of-theorem-solvability} we provide the details.  

This paper consists of four sections. Section 1 is a brief introduction to the subject and description of some applications of  Gurtin-Pipkin type equation. Section 2 contains  definitions of Sobolev space and the formulation of  correct solvability theorem. In the section 3 is provided the proof of correct solvability theorem  for the case $\xi \in (0, 1)$. Comments and obsevations of the results obtained, are given in section 4, as well as the corrections and accuracies  of the Theorem 2.2, 2.3 and Lemma 3.1 of paper \cite{RPV}. 
 
Throughout the paper, the expression $a \lesssim b$ means  $a\leq Cb$, $C>0$ and  $a \thickapprox b$ means $a \lesssim b \lesssim a$.

\section{Correct Solvability} Let $H$ be a separable Hilbert space and $A$ a self-adjoint positive operator in $H$ with a compact inverse. We associate the domain $\dom(A^{\beta})$ of the operator $A^{\beta}$, $\beta >0$, with a Hilbert space $H_{\beta}$ by introducing on $\dom(A^{\beta})$ the  norm $\|\cdot\|=\|A^{\beta}\cdot\|$, equivalent to the graph norm of the operator $A^{\beta}$. Denote by $\{e_j\}_{j=1}^{\infty}$ an othonormal basis formed by eigenvectors of $A$ corresponding to its eigenvalues $a_j$ such that $Ae_j=a_je_j, j \in \mathbb{N}$. The eigenvalues  $a_j$ are enumerated in increasing order with their multiplicity, that is, they satisfy: $0< a_1 \leq a_2 \leq \cdots \leq a_n\cdots$; where $a_n \to \infty$ as $n \to +\infty$.

By  $W_{2, \gamma}^{n}(\mathbb{R}_{+}, A^n)$ we denote the Sobolev space that consists of vector-functions on the semi-axis $\mathbb{R}_{+}=(0, \infty)$ with values in $H$ and norm  
\begin{align*}
\|u\|_{W_{2, \gamma}^{n}(\mathbb{R}_{+}, A^n)} \equiv \left(\int_{0}^{\infty}\exp(-2\gamma t)\left(\|u^{(n)}(t)\|^{2}_{H}+\|A^n u(t)\|^{2}_{H}\right) dt\right)^{1/2}, \hspace{0.3cm}\gamma \geq 0.
\end{align*}
A complete description of the space $W_{2, \gamma}^{n}(\mathbb{R}_{+}, A^n)$ and its properties are given in the monograph \cite[Chap. I]{LM}. Now, on the semi-axis $\mathbb{R}_{+}=(0, \infty)$ consider the problem
\begin{align}
\frac{d^2u}{dt^2}+A^2u- \int_{0}^{t}K(t-s)A^{2\xi} u(s) ds=f(t), \hspace{0.3cm} t \in \mathbb{R}_{+}\label{problem-valor-inicial-1},\\
u(+0)=\varphi_0, \hspace{0.4cm} u^{(1)}(+0)=\varphi_1,\hspace{0.4cm}  0<\xi<1.\label{problem-valor-inicial-2}
\end{align}
It is assumed that the vector-valued function $A^{2-\xi}f(t)$ belongs to $L_{2, \gamma_0}( \mathbb{R}_{+}, H)$ for some $\gamma_0 \geq 0$, and the scalar function $K(t)$ admits the representation 
\begin{align} \label{suma-series-exponential}
K(t)=\sum_{j=1}^{\infty}c_j\exp(-\gamma_j t),
\end{align} where $c_j>0$, $\gamma_{j+1}>\gamma_{j}>0$, $j \in \mathbb{N}$, $\gamma_j \to +\infty$ $(j \to +\infty)$ and it is assumed that 
\begin{align}\label{condition-4}
\sum_{j=1}^{\infty}\frac{c_j}{\gamma_j}<1.
\end{align} Note that if the condition (\ref{condition-4}) is satisfied, then $K \in L_1(\mathbb{R}_{+})$ and $\|K\|_{L_1} <1$. Now,  if, moreover,  we take into consideration the condition 
\begin{align}\label{condition-5}
\sum_{j=1}^{\infty} c_j<+\infty,
\end{align} then the kernel $K$ belongs to the space $W_{1}^{1}(\mathbb{R}_{+})$.
\begin{definition} A vector-valued function $u$ is called a {\em strong solution} of problem (\ref{problem-valor-inicial-1}) and (\ref{problem-valor-inicial-2}) if for some $\gamma \geq 0$, $u \in W_{2, \gamma}^{2}(\mathbb{R}_{+}, A^2)$ satisfies the equation (\ref{problem-valor-inicial-1}) almost everywhere on the semi-axis $\mathbb{R}_{+}$, as well as the initial condition (\ref{problem-valor-inicial-2}).
\end{definition} 

In the paper \cite[Theorem 1]{VR}  was shown the existence of strong solution $u$ and the well-defined solvability of system  (\ref{problem-valor-inicial-1})$-$(\ref{problem-valor-inicial-2}) for $\xi=1$. The result provided below is more general than the result obtained in \cite{VR}. But it is important to note that for $\xi=1$ we obtain the same result. 

\begin{theorem} \label{theorem-about-solvability} Suppose for some $\gamma_0 \geq 0$, $A^{2-\xi} f(t) \in L_{2, \gamma_0}(\mathbb{R}_{+}, H)$ for all $\xi \in [0,1]$. Suppose also that the condition (\ref{condition-4}) is satisfied. Then
\begin{enumerate} [\ 1)]
\item If condition (\ref{condition-5}) holds and $\varphi_0 \in H_2$, $\varphi_1 \in H_1$, $\xi \in [0,1]$ then for any $\gamma > \gamma_0$ the problems  (\ref{problem-valor-inicial-1}) and (\ref{problem-valor-inicial-2}) have a unique solution in space $W_{2, \gamma}^{2}(\mathbb{R}_{+}, A^2)$ and this solution satisfies the estimate
\begin{align}\label{solvability-solution-1}
\|u\|_{W_{2, \gamma}^{2}(\mathbb{R}_{+}, A^2)} \leq d\left(\|A^{2-\xi} f\|^{2}_{L_{2, \gamma}(\mathbb{R}_{+}, H)}+\|A^{2}\varphi_0\|_{H}+\|A\varphi_1\|_{H}\right).
\end{align}  with a constant $d$ that does not depend on the vector-valued function $f$ and the vectors $\varphi_0$, $\varphi_1$.
\item If condition (\ref{condition-5}) does not hold (that is, $K(t) \notin W_{1}^{1}(\mathbb{R}_{+})$) and $\varphi_0 \in H_{2+\xi}$, $\varphi_1 \in H_{1+\xi}$, $\xi \in (0,1]$ then for any $\gamma > \gamma_0$ the problems  (\ref{problem-valor-inicial-1}) and (\ref{problem-valor-inicial-2}) have a unique solution in space $W_{2, \gamma}^{2}(\mathbb{R}_{+}, A^2)$ and this solution satisfies the estimate
\begin{align}\label{solvability-solution-2}
\|u\|_{W_{2, \gamma}^{2}(\mathbb{R}_{+}, A^2)} \leq d\left(\|A^{2-\xi} f\|^{2}_{L_{2, \gamma}(\mathbb{R}_{+}, H)}+\|A^{2+\xi}\varphi_0\|_{H}+\|A^{1+\xi} \varphi_1\|_{H}\right).
\end{align} with a constant $d$ that does not depend on the vector-valued function $f$ and the vectors $\varphi_0$, $\varphi_1$.
\end{enumerate}
\end{theorem} 
\begin{remark} We note, moreover, that the solution $u(t) \in W_{2, \gamma}^{1}(\mathbb{R}_{+}, A^{2\xi})$ if $A^{\xi} \in L_{2, \gamma_0}(\mathbb{R}_{+}, H)$ for all  $\xi \in (0,1]$.   
\end{remark}
\section{Proof of Theorem \ref{theorem-about-solvability}} \label{proof-of-theorem-solvability}
Proof in the case of homogeneous initial conditions $\varphi_0=\varphi_1=0$. For this purpose, we need to establish well-defined solvability of Cauchy problem for hyperbolic equations on the basis of the Laplace transformation. Before proceeding, is appropriate to mention some well-known facts that will be used later. 
\begin{definition} The {\em Hardy space} $H_2(\re \zeta>\gamma, H)$ is defined as the class of functions $\hat{f}(\zeta)$ taking values in $H$, holomorphic (analytic) on the half-plane $\{\zeta \in \mathbb{C}:\re \zeta>\gamma\geq 0\}$ endowed with the norm
\begin{align}
\|\hat{f}\|^{2}_{H_2(\re \zeta>\gamma, H)}=\left(\sup_{\re \zeta>\gamma}\int_{-\infty}^{+\infty}\|\hat{f}(x+iy)\|^{2}_H\hspace{0.1cm}dy\right)^{1/2}<+\infty, \hspace{0.2cm}(\zeta=x+iy).
\end{align}Let us formulate the well-known Paley-Wiener theorem about the Hardy space  $H_2(\re \zeta>\gamma, H)$.
\end{definition}
\begin{theorem} ({\bf Paley-Wiener})
\begin{enumerate}[\ 1.]
\item The space $H_2(\re \zeta>\gamma, H)$ coincides with the set of vector-valued functions (Laplace transforms), which admit the representation
\begin{align}\label{representation-paley-wiener1}
\hat{f}(\zeta)=\frac{1}{\sqrt{2\pi}}\int_{0}^{+\infty}\exp(-\zeta t)f(t)\hspace{0.1cm}dt, 
\end{align} where $f(t) \in L_{2, \gamma_0}(\mathbb{R}_{+}, H)$, $\zeta \in \mathbb{C}$, $\re \zeta>\gamma\geq 0$
\item  For any $\hat{f}(\zeta) \in H_2(\re \zeta>\gamma, H)$ there is one and only one representation of the form (\ref{representation-paley-wiener1}), where  $f(t) \in L_{2, \gamma_0}(\mathbb{R}_{+}, H)$. Moreover, the following inversion formula holds: 
\begin{align}
f(t)=\frac{1}{\sqrt{2\pi}}\int_{-\infty}^{+\infty}\hat{f}(\gamma+iy)\exp((\gamma+iy)t)\hspace{0.1cm}dt, \hspace{0.2cm} t \in \mathbb{R}_{+}, \hspace{0.2cm}\gamma \geq 0
\end{align}
\item For  $\hat{f}(\zeta) \in H_2(\re \zeta>\gamma, H)$ and  $f(t) \in L_{2, \gamma_0}(\mathbb{R}_{+}, H)$ connected by the representation (\ref{representation-paley-wiener1}), the following relation holds: 
{\small \begin{align}
\|\hat{f}(\zeta)\|^{2}_{ H_2(\re \zeta>\gamma, H)}&\equiv\sup_{\re \zeta>\gamma}\int_{-\infty}^{+\infty}\|\hat{f}(x+iy)\|^{2}_Hdy\nonumber\\
&=\int_{0}^{+\infty}e^{-2\gamma t}\|f(t)\|^{2}_Hdt\equiv\|f(t)\|^{2}_{ L_{2, \gamma}(\mathbb{R}_{+}, H)}. 
\end{align}} 
\end{enumerate}
\end{theorem} This Theorem is well-known for scalar functions, but can be  easily extended to the case of functions with values in a separable Hilbert space (see, for instance, \cite{vvvlasov}).  
 
We begin proving the Theorem \ref{theorem-about-solvability} in the homogeneous initial conditions case $\varphi_0=\varphi_1=0$. We note that the Laplace transform $\hat{u}(\zeta)$ of any strong solution of equation (\ref{problem-valor-inicial-1}) with the initial condition (\ref{problem-valor-inicial-2}) has the form
\begin{align} \label{laplace-transfor-of-u}
\hat{u}(\zeta)=L^{-1}(\zeta)\hat{f}(\zeta), 
\end{align} where the operator-valued function $L(\zeta)$ is the symbol of equation (\ref{problem-valor-inicial-1}) and can be represented as
\begin{align}
L(\zeta)= \zeta^2I+A^2-\left(\sum_{k=1}^{\infty}\frac{c_k}{\zeta+\gamma_k}\right)A^{2\xi}, \hspace{0.4cm} 0 < \xi < 1.
\end{align} It is assumed here that there is $\gamma*\geq 0$ such that $u(t) \in  W_{2, \gamma*}^{2}(\mathbb{R}_{+}, A^2)$. This condition is necessary to apply the Laplace transform to the equation (\ref{problem-valor-inicial-1}). 

If we can prove that the vector-valued function of equation  (\ref{laplace-transfor-of-u}) is such that $A^{2}\hat{u}(\zeta)$ and $\zeta^2\hat{u}(\zeta)$ belong to the Hardy space  $H_2(\re \zeta>\gamma, H)$ for some $\gamma >\gamma_0\geq 0$, then by the Paley-Wiener theorem, we be able to prove that $A^2u(t)$ and $d^2u(t)/dt^2$ belong to $L_{2, \gamma}(\mathbb{R}_{+}, H)$ and, therefore we will have shown that $u(t) \in W_{2, \gamma}^{2}(\mathbb{R}_{+}, A^2)$. That is, the solvability of system  (\ref{problem-valor-inicial-1})$-$(\ref{problem-valor-inicial-2}) in the space  $W_{2, \gamma}^{2}(\mathbb{R}_{+}, A^2)$ will be established. 

With this in mind, let us consider the projection $\hat{u}_n(\zeta)$ of the vector-valued function $\hat{u}(\zeta)$ to the one-dimensional subspace spanned by the vector $e_n$:
\begin{align} \label{laplace-transfor-of-u_n}
\hat{u}_n(\zeta)=\ell_{n}^{-1}(\zeta)\hat{f}_n(\zeta), 
\end{align} where $\hat{f}_n(\zeta)=(\hat{f}(\zeta), e_n)$ and 
\begin{align*}
\ell_n(\zeta):=(L(\zeta)e_n, e_n)=\zeta^2+a^{2}_n-\left(\sum_{k=1}^{\infty}\frac{c_k}{\zeta+\gamma_k}\right)a_{n}^{2\xi}
\end{align*}
 The restriction of  $A^{2}\hat{u}(\zeta)$ to the one-dimensional space spanned by $e_n$ has the form
\begin{align} 
\left(A^2\hat{u}(\zeta)e_n, e_n	\right)= \frac{a^{\xi}_n\hat{g}_n(\zeta)}{\ell_n(\zeta)}, \hspace{0.5cm} 0 \leq \xi \leq 1
\end{align} where $\hat{g}_n(\zeta)$ is the n-th coordinate of the vector-valued function $\hat{g}(\zeta)=A^{2-\xi}\hat{f}(\zeta)$. According to the conditions of the Theorem \ref{theorem-about-solvability}, the vector-valued function  $g(t)=A^{2-\xi}f(t)$ belongs to the space  $L_{2, \gamma_0}(\mathbb{R}_{+}, H)$, and therefore, its Laplace transform  $\hat{g}(\zeta)=A^{2-\xi}\hat{f}(\zeta)$ belongs to Hardy space $H_2(\re \zeta>\gamma_0, H)$.

In order to prove that $A^{2}\hat{u}(\zeta)$ belongs to $H_2(\re \zeta>\gamma, H)$, it is suffices to establish the estimate
\begin{align}
\sup_{\substack{\re \zeta>\gamma \\ n \in \mathbb{N} }}\left|\frac{a^{\xi}_n}{\ell_n(\zeta)}\right|\leq const, \hspace{0.4cm} \mbox{for all $\xi \in [0, 1]$}
\end{align} which is uniform with respect to $ \zeta (\re \zeta>\gamma)$ and $ n \in \mathbb{N}$.  

For this purpose, consider the function $\mathfrak{m}_n(\zeta)=\frac{\ell_n(\zeta)}{a^{2}_n}$. Let us estimate  $\mathfrak{m}_n(\zeta)$ from below by means of its real and imaginary parts:
\begin{align*}
\re \mathfrak{m}_n(\zeta)&=\frac{x^{2}-y^{2}}{a^{2}_n}+1-\frac{1}{a^{2(1-\xi)}_n}\left(\sum_{k=1}^{\infty}\frac{c_k(x+\gamma_k)}{(x+\gamma_k)^{2}+y^{2}}\right), \hspace{0.2cm} \zeta=x+iy\\
\im \mathfrak{m}_n(\zeta)&=\frac{2xy}{a^{2}_n}+\frac{y}{a^{2(1-\xi)}_n}\left(\sum_{k=1}^{\infty}\frac{c_k}{(x+\gamma_k)^{2}+y^{2}}\right).
\end{align*} 

First, we find a lower bound for $\left|\im \mathfrak{m}_n(\zeta)\right|$ for $|y|>x$, where $x>\gamma>\gamma_1\geq 0$:
\begin{align*}
\left|\im \mathfrak{m}_n(\zeta)\right|>\frac{2x|y|}{a^{2}_n}+\frac{1}{|y|a^{2(1-\xi)}_n}\left(\sum_{k=1}^{\infty}\frac{c_k}{(1+\frac{\gamma_k}{|y|})^{2}+1}\right)>\frac{2\gamma y^{2}+k_0(\gamma) a^{2\xi}_n}{|y|a^{2}_n}, 
\end{align*}where $k_0(\gamma)=\frac{c_1}{(1+\frac{\gamma_1}{\gamma})^{2}+1}$. Hence, for $|y|>x$ with $x>\gamma>\gamma_1\geq 0$ we have
\begin{align} \label{estimate1}
\frac{1}{|\ell_n(\zeta)|}=\frac{1}{a^{2}_n|\im \mathfrak{m}_n(\zeta)|}<\frac{|y|}{2\gamma y^{2}+k_0(\gamma) a^{2\xi}_n}<\frac{1}{ a^{\xi}_n\sqrt{2 \gamma \cdot k_0(\gamma)}}. 
\end{align} 

Second, we estimate $\left|\re \mathfrak{m}_n(\zeta)\right|$ from below for $|y|<x$, where $x>\gamma>\gamma_1\geq 0$. Note that
\begin{align*}
\frac{1}{a^{2(1-\xi)}_n}\left(\sum_{k=1}^{\infty}\frac{c_k(x+\gamma_k)}{(x+\gamma_k)^{2}+y^{2}}\right)<\frac{1}{a^{2(1-\xi)}_n}\left(\sum_{k=1}^{\infty}\frac{c_k}{x+\gamma_k}\right)<\frac{1}{a^{2(1-\xi)}_n}\left(\sum_{k=1}^{\infty}\frac{c_k}{\gamma_k}\right)<1. 
\end{align*} 
It follows that 
\begin{align*}
\left|\re \mathfrak{m}_n(\zeta)\right| > \left|1-\frac{1}{a^{2(1-\xi)}_n}\left(\sum_{k=1}^{\infty}\frac{c_k}{\gamma_k}\right)\right|>0. 
\end{align*} 
Therefore,  for $|y|<x$, with $x>\gamma>\gamma_1\geq 0$, we have
\begin{align}\label{estimate2}
\frac{1}{|\ell_n(\zeta)|}<\frac{1}{a^{2}_n\left|\re \mathfrak{m}_n(\zeta)\right|}<\frac{1}{a^{2}_n \left|1-\frac{1}{a^{2(1-\xi)}_n}\sum_{k=1}^{\infty}\frac{c_k}{\gamma_k}\right|}. 
\end{align} From the estimates (\ref{estimate1}) and (\ref{estimate2}) we obtain

\begin{align*}
\left|\frac{a^{\xi}_n}{\ell_n(\zeta)}\right|&<\frac{1}{\sqrt{2 \gamma \cdot k_0(\gamma)}}\\
\left|\frac{a^{\xi}_n}{\ell_n(\zeta)}\right|&<\frac{1}{a^{2-\xi}_n \left|1-\frac{1}{a^{2(1-\xi)}_n}\sum_{k=1}^{\infty}\frac{c_k}{\gamma_k}\right|}< \frac{1}{a^{2-\xi}_1 \left|1-\frac{1}{a^{2(1-\xi)}_1}\sum_{k=1}^{\infty}\frac{c_k}{\gamma_k}\right|}. 
\end{align*}
Therefore, 
\begin{align}\label{minimum-estimates}
\sup_{\substack{\re \zeta>\gamma \\ n \in \mathbb{N} }}\left|\frac{a^{\xi}_n}{\ell_n(\zeta)}\right|< \frac{1}{\min \left(\sqrt{2 \gamma \cdot k_0(\gamma)},  a^{2-\xi}_1 \left|1-\frac{1}{a^{2(1-\xi)}_1}\sum_{k=1}^{\infty}\frac{c_k}{\gamma_k}\right|\right)}, \hspace{0.2cm} \mbox{for all $\xi \in [0, 1]$}.
\end{align}
\begin{remark}\label{remark-symbol-less-const} The estimate (\ref{minimum-estimates}) implies that 
\begin{align}
\sup_{\substack{\re \zeta>\gamma}}\left\|A^{\xi}L^{-1}(\zeta)\right\|\leq const.
\end{align}
\end{remark}
The Hardy space $H_2(\re \zeta>\gamma, H)$ is invariant with respect to multiplication of functions of the form $\frac{a^{\xi}_n}{\ell_n(\zeta)}$, since they are analytic and bounded in view of  (\ref{minimum-estimates}). Therefore, the inclusion  $\hat{g}(\zeta)=A^{2-\xi}\hat{f}(\zeta) \in H_2(\re \zeta>\gamma, H)$ implies that $A^{2}\hat{u}(\zeta)$  belongs to $H_2(\re \zeta>\gamma, H)$. 

Let us establish the estimate for the norm of vector-valued function  $A^{2}u(t) \in L_{2, \gamma}(\mathbb{R}_{+}, H)$. From (\ref{laplace-transfor-of-u}), it follows that 
\begin{align} 
A^{2}\hat{u}(\zeta)=A^{2}L^{-1}(\zeta)\hat{f}(\zeta)=A^{\xi}L^{-1}(\zeta)A^{2-\xi}\hat{f}(\zeta). 
\end{align} 
This function can be represented in the form 
\begin{align*} 
A^{2}\hat{u}(\zeta)=\sum_{k=1}^{\infty}\frac{a^{\xi}_k}{\ell_k(\zeta)}\cdot a_k^{2-\xi}\hat{f}_k(\zeta)e_k. 
\end{align*} 
According to the hypothesis of Theorem \ref{theorem-about-solvability}, the vector-valued function  $A^{2-\xi}f(t) \in L_{2, \gamma_0}(\mathbb{R}_{+}, H)$. Therefore, by the Paley-Wiener theorem,  $A^{2-\xi}\hat{f}(\zeta) \in H_2(\re \zeta>\gamma_0, H)$ and
\begin{align*} 
\|A^{2-\xi}f\|_{L_{2, \gamma_0}(\mathbb{R}_{+}, H)}=\|A^{2-\xi}\hat{f}\|_{ H_2(\re \zeta>\gamma_0, H)}.
\end{align*}

By (\ref{minimum-estimates}) and Paley-Wiener theorem, the following relations  hold: 
\begin{align*} 
\|A^{2}u(t)\|^{2}_{L_{2, \gamma}(\mathbb{R}_{+}, H)}=\|A^{2}\hat{u}(\zeta)\|^{2}_{ H_2(\re \zeta>\gamma, H)}=\|A^{\xi}L^{-1}(\zeta)A^{2-\xi}\hat{f}(\zeta)\|^{2}_{ H_2(\re \zeta>\gamma, H)}
\end{align*} But
{\small \begin{align} 
\|A^{\xi}L^{-1}(\zeta)A^{2-\xi}\hat{f}(\zeta)\|^{2}_{ H_2(\re \zeta>\gamma, H)}&=\sup_{\substack{\re \zeta>\gamma}}\int_{-\infty}^{+\infty}\left(\sum_{k=1}^{\infty}\left|\frac{a^{\xi}_k}{\ell_k(\zeta)}\cdot a_k^{2-\xi}\hat{f}_k(\zeta)\right|^{2}\right)dy\nonumber\\
&\leq \sup_{\substack{\re \zeta>\gamma\\ k \in \mathbb{N} }} \left|\frac{a^{\xi}_k}{\ell_k(\zeta)}\right|^{2}\cdot \sup_{\substack{\re \zeta>\gamma}}\int_{-\infty}^{+\infty}\left(\sum_{k=1}^{\infty}\left|a_k^{2-\xi}\hat{f}_k(\zeta)\right|^{2}\right)dy\\
&\leq \sup_{\substack{\re \zeta>\gamma\\ k \in \mathbb{N} }} \left|\frac{a^{\xi}_k}{\ell_k(\zeta)}\right|^{2} \|A^{2-\xi}\hat{f}(\zeta)\|^{2}_{H_2(\re \zeta>\gamma, H)}\leq d^{2}_1 \|A^{2-\xi}f\|^{2}_{L_{2, \gamma}(\mathbb{R}_{+}, H)}.\nonumber
\end{align}} where $d_1= \sup_{\substack{\re \zeta>\gamma\\ k \in \mathbb{N} }} \left|\frac{a^{\xi}_k}{\ell_k(\zeta)}\right|$. Hence, we obtain the inclusion $A^{2}u(t) \in L_{2, \gamma}(\mathbb{R}_{+}, H)$ and the estimate 
\begin{align}\label{estimate-d-1} 
\|A^{2}u\|_{L_{2, \gamma}(\mathbb{R}_{+}, H)}\leq d_1\left(\|A^{2-\xi}f\|_{L_{2, \gamma}(\mathbb{R}_{+}, H)}\right).
\end{align} is valid. 

Now let us prove that $\zeta^2\hat{u}$ also belongs to $H_2(\re \zeta>\gamma, H)$ . Set 
\begin{align*}
\Psi (\zeta):= \sum_{k=1}^{\infty}\frac{c_k}{\zeta +\gamma_k}.
\end{align*}
Note that for $\re \zeta>\gamma$ we can write 
\begin{align*} 
I= \zeta^{2}L^{-1}(\zeta)+\left(1-\Psi(\zeta)A^{-2(1-\xi)}\right) A^{2}L^{-1}(\zeta). 
\end{align*}
Hence, for $\re \zeta>\gamma$, we obtain
\begin{align} \label{function-Psi=suma}
\hat{f}(\zeta)&= \zeta^{2}L^{-1}(\zeta)\hat{f}(\zeta)+\left(1-\Psi(\zeta)A^{-2(1-\xi)}\right) A^{2}L^{-1}(\zeta)\hat{f}(\zeta)\nonumber\\
&= \zeta^{2}\hat{u}(\zeta)+\left(1-\Psi(\zeta) A^{-2(1-\xi)}\right) A^{\xi}L^{-1}(\zeta)A^{2-\xi}\hat{f}(\zeta).  
\end{align} It is our understanding that $A^{2-\xi}\hat{f}(\zeta) \in H_2(\re \zeta>\gamma, H)$ and  $\sup_{\substack{\re \zeta>\gamma}}\left\|A^{\xi}L^{-1}(\zeta)\right\|\leq const.$ From (\ref{function-Psi=suma}), it follows that 
\begin{align} 
\zeta^{2}\hat{u}(\zeta)= \hat{f}(\zeta)-\left(1-\Psi(\zeta)A^{-2(1-\xi)}\right) A^{\xi}L^{-1}(\zeta)A^{2-\xi}\hat{f}(\zeta).  
\end{align} Therefore, the function $\zeta^{2}\hat{u}_n(\zeta)$ can be represented in the form
\begin{align} 
\zeta^{2}\hat{u}_n(\zeta)= \hat{f}_n(\zeta)-\left(1-\frac{\Psi(\zeta)}{a^{2(1-\xi)}_n}\right) \frac{a^{\xi}_n}{\ell_n(\zeta)} \hat{g}_n(\zeta) 
\end{align}
Under the assumptions imposed on the sequences $\{c_k\}_{k=1}^{\infty}$ and $\{\gamma_k\}_{k=1}^{\infty}$, the function $\Psi (\zeta)$ is analytic and bounded on the half-plane $\{\zeta: \re \zeta>\gamma\}$. Indeed,
{\small \begin{align} 
\left|1-\frac{\Psi (\zeta)}{a^{2(1-\xi)}_n}\right|\leq 1+\left|\frac{\Psi (\zeta)}{a^{2(1-\xi)}_n}\right|&<1+\frac{1}{a^{2(1-\xi)}_n}\sum_{k=1}^{\infty}\frac{c_k}{\gamma_k\left|\frac{\zeta}{\gamma_k} +1\right|}\nonumber\\
&<1+\frac{1}{a^{2(1-\xi)}_n}\sum_{k=1}^{\infty}\frac{c_k}{\gamma_k\left(\frac{\re \zeta}{\gamma_k} +1\right)}<1+\frac{1}{a^{2(1-\xi)}_n}\sum_{k=1}^{\infty}\frac{c_k}{\gamma_k\left(\frac{\gamma}{\gamma_k} +1\right)}\nonumber\\
&<1+\frac{1}{a^{2(1-\xi)}_n}\sum_{k=1}^{\infty}\frac{c_k}{\gamma_k}<2. 
\end{align}}
Since the vector-function $A^{2-\xi}f (t)\in L_{2, \gamma}(\mathbb{R}_{+}, H) $  then its Laplace transform $A^{2-\xi}\hat{f} (\zeta)$ belongs to Hardy space $  H_2(\re \zeta>\gamma, H)$, that is, 
\begin{align*} 
\|A^{2-\xi}\hat{f}\|^{2}_{ H_2(\re \zeta>\gamma, H)}=\sup_{\substack{\re \zeta>\gamma}}\int_{-\infty}^{+\infty}\|A^{2-\xi}\hat{f}(x+iy)\|^{2}_{ H}dy<+\infty, \hspace{0.2cm} \zeta=x+iy
\end{align*}
Hence we obtain the following relation
\begin{align} \label{function-f-also-in-H-2}
\|A^{2-\xi}\hat{f}\|^{2}_{ H_2(\re \zeta>\gamma, H)}&=\sup_{\substack{\re \zeta>\gamma \\ n \in \mathbb{N}}}\int_{-\infty}^{+\infty}\left(\sum_{n=1}^{\infty}|a^{2-\xi}_n\hat{f}_n(x+iy)|^{2}\right)dy\nonumber\\
&>\sup_{\substack{\re \zeta>\gamma \\ n \in \mathbb{N}}}a^{2(2-\xi)}_1\int_{-\infty}^{+\infty}\left(\sum_{n=1}^{\infty}|\hat{f}_n(x+iy)|^{2}\right)dy\nonumber\\
&=a^{2(2-\xi)}_1\sup_{\substack{\re \zeta>\gamma}}\int_{-\infty}^{+\infty}\|\hat{f}(x+iy)\|^{2}_{ H}dy=a^{2(2-\xi)}_1\|\hat{f}(\zeta)\|^{2}_{ H_2(\re \zeta>\gamma, H)}.
\end{align}
 Therefore,  $\hat{f}(\zeta) \in H_2(\re \zeta>\gamma, H)$ and  $f(t) \in L_{2, \gamma}(\mathbb{R}_{+}, H)$. 
Now, taking into account (\ref{minimum-estimates}) and  (\ref{function-f-also-in-H-2}) we obtain the following estimate: 
{\small \begin{align*} 
\sup_{\substack{\re \zeta>\gamma \\ n \in \mathbb{N}}}\int_{-\infty}^{+\infty}|\zeta^{2}\hat{u}_n(\zeta)|^{2}dy< \sup_{\substack{\re \zeta>\gamma\\ n \in \mathbb{N}}}\left(\frac{1}{a^{2(2-\xi)}_1} +4\left|\frac{a^{\xi}_n}{\ell_n(\zeta)}\right|^2\right)\cdot \sup_{\substack{\re \zeta>\gamma\\ n \in \mathbb{N}}}\int_{-\infty}^{+\infty} |\hat{g}_n(\zeta)|^{2}dy<+\infty.
\end{align*}}
Therefore $\zeta^{2}\hat{u}_n(\zeta) \in  H_2(\re \zeta>\gamma, \mathbb{C})$ and  $\frac{d^{2}}{dt^{2}}u_n(t) \in L_{2, \gamma}(\mathbb{R}_{+}, \mathbb{C})$. 

By (\ref{minimum-estimates}) and Paley-Wiener theorem we have: 
\begin{align} 
\left\|\frac{d^{2}}{dt^{2}}u(t)\right\|^{2}_{L_{2, \gamma}(\mathbb{R}_{+}, H)}&=\|\zeta^{2}\hat{u}(\zeta)\|^{2}_{ H_2(\re \zeta>\gamma, H)}\nonumber\\
&=\| \hat{f}(\zeta)-\left(1-\Psi(\zeta)A^{-2(1-\xi)}\right) A^{\xi}L^{-1}(\zeta)A^{2-\xi}\hat{f}(\zeta)\|^{2}_{ H_2(\re \zeta>\gamma, H)}\nonumber\\
&<\frac{1}{a^{2(2-\xi)}_1}\|A^{2-\xi}\hat{f}(\zeta)\|^{2}_{ H_2(\re \zeta>\gamma, H)} +4\|A^{\xi}L^{-1}(\zeta)A^{2-\xi}\hat{f}(\zeta)\|^{2}_{ H_2(\re \zeta>\gamma, H)}\nonumber\\
&\leq \frac{1}{a^{2(2-\xi)}_1}\|A^{2-\xi}\hat{f}(\zeta)\|^{2}_{ H_2(\re \zeta>\gamma, H)} +4\sup_{\substack{\re \zeta>\gamma\\ k \in \mathbb{N} }} \left|\frac{a^{\xi}_k}{\ell_k(\zeta)}\right|^{2} \|A^{2-\xi}\hat{f}(\zeta)\|^{2}_{H_2(\re \zeta>\gamma, H)}\nonumber\\
&< \left(\frac{1}{a^{2-\xi}_1}+2\sup_{\substack{\re \zeta>\gamma\\ k \in \mathbb{N} }} \left|\frac{a^{\xi}_k}{\ell_k(\zeta)}\right|\right)^{2} \|A^{2-\xi}\hat{f}(\zeta)\|^{2}_{H_2(\re \zeta>\gamma, H)}\nonumber\\
&\leq d^{2}_2 \|A^{2-\xi}f\|^{2}_{L_{2, \gamma}(\mathbb{R}_{+}, H)},
\end{align} where $d_2=\left(\frac{1}{a^{2-\xi}_1}+2d_1\right)$. Hence, we obtain the inclusion $\frac{d^{2}}{dt^{2}}u(t) \in L_{2, \gamma}(\mathbb{R}_{+}, H)$ and the estimate 
\begin{align}\label{estimate-d-2}
\left\|\frac{d^{2}}{dt^{2}}u(t)\right\|_{L_{2, \gamma}(\mathbb{R}_{+}, H)}\leq d_2\left(\|A^{2-\xi}f\|_{L_{2, \gamma}(\mathbb{R}_{+}, H)}\right)
\end{align} is valid. Accordingly, combining the estimates (\ref{estimate-d-1}) and (\ref{estimate-d-2}), we come to the desired inequality
\begin{align}\label{desired-inequality}
\|u(t)\|_{W_{2, \gamma}^2(\mathbb{R}_{+}, A^2)}\leq d\left(\|A^{2-\xi}f\|_{L_{2, \gamma}(\mathbb{R}_{+}, H)}\right).
\end{align} with a constant $d$ independent of $f$. Implying that the equation (\ref{problem-valor-inicial-1}) has a solution $u(t)$, which belongs to space $W_{2, \gamma}^2(\mathbb{R}_{+}, A^2)$ and the estimate (\ref{desired-inequality}) holds. 
 
Now, let us prove that the solution $u(t)$  the initial conditions $u(+0)=0$ and $u^{(1)}(+0)=0$. 
\begin{remark}\label{remark-in-hardy-space} If $\varphi(\zeta) \in H_2(\re \zeta>\gamma, \mathbb{C})$, then for any $\eta>\gamma$ there is a sequence $\{\eta_k\}^{\infty}_{k=1}$ such that $\lim_{k \to \infty}\eta_k=+\infty$ and 
\begin{align*}
\lim_{k \to \infty}\int_{\gamma}^{\eta}|\varphi(x\pm i\eta_k)|^2dx=0. 
\end{align*} Indeed, for any $\eta>\gamma$ and $\eta_k>0$, we have
\begin{align*}
\int_{-\eta_k}^{\eta_k}\left(\int_{\gamma}^{\eta}|\varphi(x\pm i\eta_k)|^2dx\right)dy \leq \int_{\gamma}^{\eta}\left(\int_{-\infty}^{+\infty}|\varphi(x\pm i\eta_k)|^2dy\right) dx<+\infty.
\end{align*} 
Therefore, for any $\eta>0$ there is a sequence $\{\eta_k\}^{\infty}_{k=1}$ such that $\lim_{k \to \infty}\eta_k=+\infty$ and 
\begin{align*}
\lim_{k \to \infty}\int_{\gamma}^{\eta}|\varphi(x\pm i\eta_k)|^2dx=0. 
\end{align*}
\end{remark} Now it remains to use the Cauchy inequality.

The above reasoning shows that $u(t) \in L_{2, \gamma}(\mathbb{R}_{+}, H)$, and therefore, $\hat{u}(\zeta) \in H_2(\re \zeta>\gamma, H)$ and $\hat{u}_n(\zeta) \in H_2(\re \zeta>\gamma, \mathbb{C})$. Moreover, let us prove that $\zeta \hat{u}_n(\zeta)\in H_2(\re \zeta>\gamma, \mathbb{C})$. Indeed, 
{\small \begin{align*}
\sup_{\substack{\re \zeta>\gamma \\ n \in \mathbb{N}}}\int_{-\infty}^{+\infty}|(\re \zeta+iy)\hat{u}_n(\re \zeta+iy)|^{2}dy&=\sup_{\substack{\re \zeta>\gamma \\ n \in \mathbb{N}}}\int_{-\infty}^{+\infty}\frac{|(\re \zeta+iy)^2\hat{u}_n(\re \zeta+iy)|^{2}}{(\re \zeta)^2+y^2}dy\\
&<\frac{1}{\gamma}\sup_{\substack{\re \zeta>\gamma \\ n \in \mathbb{N}}}\int_{-\infty}^{+\infty}|\zeta^2\hat{u}_n(\zeta)|^{2}dy<+\infty, 
\end{align*} } since $\zeta^2 \hat{u}_n(\zeta)\in H_2(\re \zeta>\gamma, \mathbb{C})$. By the Paley-Wiener theorem, we obtain
{\small \begin{align*}
\hat{u}_n(+0)=\frac{1}{\sqrt{2\pi}}\lim_{\eta_k \to \infty} \int_{-\eta_k}^{\eta_k}\hat{u}_n(x+iy)dy=\frac{1}{\sqrt{2\pi i}}\lim_{\eta_k \to \infty} \int_{\gamma-i\eta_k}^{\gamma+i\eta_k}\hat{u}_n(\zeta)d\zeta\\
\hat{u}^{(1)}_n(+0)= \frac{1}{\sqrt{2\pi}}\lim_{\eta_k \to \infty} \int_{-\eta_k}^{\eta_k}(x+iy)\hat{u}_n(x+iy)dy=\frac{1}{\sqrt{2\pi i}}\lim_{\eta_k \to \infty} \int_{\gamma-i\eta_k}^{\gamma+i\eta_k}\zeta\hat{u}_n(\zeta)d\zeta. 
\end{align*}}
The functions $\hat{u}_n(+0)$ and $\hat{u}^{(1)}_n(+0)$ are analytics on the right half-plane $\re \zeta>\gamma\geq 0$, and therefore, by the Cauchy theorem, for any $\eta>\gamma$, we have 
{\small \begin{align*}
\int_{\gamma-i\eta_k}^{\gamma+i\eta_k}\hat{u}_n(\zeta)d\zeta&=  \left(\int_{\gamma-i\eta_k}^{\eta-i\eta_k}-\int_{\gamma+i\eta_k}^{\eta+i\eta_k}+\int_{\eta-i\eta_k}^{\eta+i\eta_k}\right)\hat{u}_n(\zeta)d\zeta\\
&=\int_{\gamma}^{\eta}\hat{u}_n(x-i\eta_k)dx-\int_{\gamma}^{\eta}\hat{u}_n(x+i\eta_k)dx+i\int_{-\eta_k}^{\eta_k}\hat{u}_n(\eta+i y)dy.
\end{align*}} and 
{\small \begin{align*}
\int_{\gamma-i\eta_k}^{\gamma+i\eta_k}\zeta\hat{u}_n(\zeta)d\zeta&=  \left(\int_{\gamma-i\eta_k}^{\eta-i\eta_k}-\int_{\gamma+i\eta_k}^{\eta+i\eta_k}+\int_{\eta-i\eta_k}^{\eta+i\eta_k}\right)\zeta\hat{u}_n(\zeta)d\zeta\\
&=\int_{\gamma}^{\eta}(x-i\eta_k)\hat{u}_n(x-i\eta_k)dx-\int_{\gamma}^{\eta}(x+i\eta_k)\hat{u}_n(x+i\eta_k)dx+\\
&\hspace{5cm}+i\int_{-\eta_k}^{\eta_k}(\eta+i y)\hat{u}_n(\eta+i y)dy.
\end{align*}}

According to Remark \ref{remark-in-hardy-space}, we have
\begin{align*}
&\lim_{k \to \infty}\int_{\gamma}^{\eta}|\hat{u}_n(x\pm i\eta_k)|^2dx=0\\
&\lim_{k \to \infty}\int_{\gamma}^{\eta}|(x\pm i\eta_k)\hat{u}_n(x\pm i\eta_k)|^2dx=0, 
\end{align*} since $\zeta^2\hat{u}_n(\zeta) \in H_2(\re \zeta>\gamma, \mathbb{C})$. Therefore, for $\eta>\gamma$, we find that
{\small \begin{align*}
|\hat{u}_n(+0)|&\leq \frac{1}{\sqrt{2\pi}}\lim_{\eta_k \to \infty} \int_{-\eta_k}^{\eta_k}|\hat{u}_n(\eta+i y)|dy=\frac{1}{\sqrt{2\pi}} \int_{-\infty}^{+\infty}\left|\frac{(\eta+iy)^2\hat{u}_n(\eta+iy)}{(\eta+iy)^2}\right|dy\\
&\leq \frac{1}{\sqrt{2\pi}} \left(\int_{-\infty}^{+\infty}|(\eta+iy)^2\hat{u}_n(\eta+iy)|^2dy\right)^{1/2} \left(\int_{-\infty}^{+\infty}\frac{dy}{(\eta^2+y^2)^2}\right)^{1/2}\\
&\lesssim \frac{1}{\eta^{3/2}}
\end{align*}}
{\small \begin{align*}
|\hat{u}^{(1)}_n(+0)|&\leq \frac{1}{\sqrt{2\pi}}\lim_{\eta_k \to \infty} \int_{-\eta_k}^{\eta_k}|(\eta+i y)\hat{u}_n(\eta+i y)|dy=\frac{1}{\sqrt{2\pi}}\int_{-\infty}^{+\infty}\left|\frac{(\eta+iy)^2\hat{u}_n(\eta+iy)}{\eta+iy}\right|dy\\
&\leq \frac{1}{\sqrt{2\pi}} \left(\int_{-\infty}^{+\infty}|(\eta+iy)^2\hat{u}_n(\eta+iy)|^2dy\right)^{1/2} \left(\int_{-\infty}^{+\infty}\frac{dy}{\eta^2+y^2}\right)^{1/2}\\
&\lesssim \frac{1}{\eta^{1/2}}
 \end{align*}}
It follows that for $\eta \to +\infty$,  $u(+0)=0$ and $u^{(1)}(+0)=0$. 

Finally, let us prove that the solution $u(t)$ satisfies the equation (\ref{problem-valor-inicial-1}). By Paley-Wiener theorem, we have
{\small \begin{align*}
u(t)= \frac{1}{\sqrt{2\pi}}\lim_{\eta \to +\infty} \int_{-\eta}^{\eta}L^{-1}(\gamma+i y)\hat{f}(\gamma+i y)\mathrm{e}^{(\gamma+i y)t}dy=\frac{1}{\sqrt{2\pi i}}\lim_{\eta \to +\infty} \int_{\gamma-i\eta}^{\gamma+i\eta}L^{-1}(\zeta)\hat{f}(\zeta)\mathrm{e}^{\zeta t}d\zeta.
 \end{align*}} Hence, we get
 \begin{align}
\frac{d^2}{dt^2}u(t)=\frac{1}{\sqrt{2\pi i}}\lim_{\eta \to +\infty} \int_{\gamma-i\eta}^{\gamma+i\eta}\zeta^2L^{-1}(\zeta)\hat{f}(\zeta)\mathrm{e}^{\zeta t}d\zeta \label{u(t)-satis-equation1-1}\\
A^2u(t)=\frac{1}{\sqrt{2\pi i}}\lim_{\eta \to +\infty} \int_{\gamma-i\eta}^{\gamma+i\eta}A^2L^{-1}(\zeta)\hat{f}(\zeta)\mathrm{e}^{\zeta t}d\zeta \label{u(t)-satis-equation1-2}
 \end{align}
{\small \begin{align}
\int_{0}^{t}K(t-s)A^{2\xi}u(s)ds&=\frac{1}{\sqrt{2\pi i}}\lim_{\eta \to +\infty} \int_{0}^{t}\sum_{k=1}^{\infty}c_k \mathrm{e}^{-\gamma_k(t-s)}\left(\int_{\gamma-i\eta}^{\gamma+i\eta}A^{2\xi}L^{-1}(\zeta)\hat{f}(\zeta)\mathrm{e}^{\zeta t}d\zeta\right)ds\nonumber\\
&=\frac{1}{\sqrt{2\pi i}}\lim_{\eta \to +\infty}\int_{\gamma-i\eta}^{\gamma+i\eta} A^{2\xi}L^{-1}(\zeta)\hat{f}(\zeta) \left(\sum_{k=1}^{\infty}c_k \int_{0}^{t}\mathrm{e}^{-\gamma_k(t-s)} \mathrm{e}^{\zeta t} ds\right)d\zeta\nonumber\\
&=\frac{1}{\sqrt{2\pi i}}\lim_{\eta \to +\infty}\int_{\gamma-i\eta}^{\gamma+i\eta} \left(\sum_{k=1}^{\infty}\frac{c_k}{\zeta+\gamma_k}\right)A^{2\xi}L^{-1}(\zeta)\hat{f}(\zeta)\mathrm{e}^{\zeta t} d\zeta. \label{u(t)-satis-equation1-3}
 \end{align}} From (\ref{u(t)-satis-equation1-1})-(\ref{u(t)-satis-equation1-3}), it follows that $u(t)$ satisfies the equation (\ref{problem-valor-inicial-1}). 
 
 Let us turn to the proof of Theorem \ref{theorem-about-solvability} in the case of nonhomogeneous initial conditions. In the system (\ref{problem-valor-inicial-1})-(\ref{problem-valor-inicial-2}), set 
\begin{align*} 
u(t)= \cos (At)\varphi_0+A^{-1}\sin (At)\varphi_1+\omega(t). 
\end{align*} Then for the function $\omega(t)$ we obtain the system

\begin{align*}
\frac{d^2}{dt^2}\omega(t)+&A^2 \omega(t)- \int_{0}^{t}K(t-s)A^{2\xi}ds=f_1(t), \hspace{0.2cm}t \in \mathbb{R}_+ \\
&\omega(+0)=\omega^{(1)}(+0)=0,
\end{align*}where $f_1(t)=f(t)-h(t)$ and 
\begin{align}\label{expression-function-h} 
h(t)&=\int_{0}^{t}K(t-s)A^{2\xi}\left(\cos (At)\varphi_0+A^{-1}\sin (At)\varphi_1\right)ds\\
A^{2-\xi}h(t)&=\int_{0}^{t}K(t-s)\left(A^{2+\xi}\cos (At)\varphi_0+A^{1+\xi}\sin (At)\varphi_1\right)ds
\end{align} Let us prove that the vector-valued function $f_1(t)$ satisfies the conditions of Theorem \ref{theorem-about-solvability} with homogeneous initial conditions. Indeed, 
\begin{align*}
\|A^{2-\xi}f_1(t)\|_{L_{2, \gamma} (\mathbb{R}_+ ,H)}\leq\|A^{2-\xi}f(t)\|_{L_{2, \gamma} (\mathbb{R}_+ ,H)}+\|A^{2-\xi}h(t)\|_{L_{2, \gamma} (\mathbb{R}_+ ,H)}.
\end{align*} Let us estimate the norm $\|A^{2-\xi}h(t)\|_{L_{2, \gamma} (\mathbb{R}_+ ,H)}$.

1. Suppose that the condition (\ref{condition-5}) holds. The direct integration shows that
\begin{align}
\int_{0}^{t}\mathrm{e}^{-\gamma_k(t-s)} \cos (As)ds=\left(A^2+\gamma^2_k\right)^{-1}\left\{\gamma_k\{\cos (At)-\exp (-\gamma_kt)\}+A\sin (At)\right\}\label{integral-cos}\\
\int_{0}^{t}\mathrm{e}^{-\gamma_k(t-s)}\sin (As)ds=  \left(A^2+\gamma^2_k\right)^{-1}\left\{A\{\exp (-\gamma_kt)-\cos (At)\}+\gamma_k\sin (At)\right\} \label{integral-sin}
\end{align} In what follows, we need the following argument.
\begin{remark} The following inequality holds 
\begin{align} \label{inequality-gamma}
\|\left(A^2+\gamma^2_k\right)^{-1}\|^{2}_{H} \lesssim  \gamma^{-2}_k\|A^{-1}\|^{2}_{H}.
\end{align}Indeed, note that for any vector $v \in H$, $\|v\|_{H}=1$, the following relations are valid: 
\begin{align*}
\|\left(A^2+\gamma^2_k\right)^{-1}\|^{2}_{H}= \sum_{n=1}^{\infty}\left(a^2_n+\gamma^2_n\right)^{-2}|v_n|^2 \lesssim \sum_{n=1}^{\infty}(a_n\gamma_n)^{-2}|v_n|^2 = \gamma^{-2}_k\|A^{-1}v\|^{2}_{H},
\end{align*}where $v_n=(v, e_n)$. This implies (\ref{inequality-gamma}), since operator $A$ is self-adjoint and $\gamma_k$ is positive.
\end{remark} Now, using (\ref{expression-function-h}), (\ref{integral-cos}), (\ref{integral-sin}), the inequalities $\|\cos (At)\|\leq 1$, $\|\sin (At)\|\leq 1$ and the above remark, we obtain the following result: 
 {\small \begin{align*}
\|A^{2-\xi}h(t)\|_{L_{2, \gamma} (\mathbb{R}_+ ,H)}&=\left\|\sum_{k=1}^{\infty}c_k\int_{0}^{t}\mathrm{e}^{-\gamma_k(t-s)}\left\{A^{2+\xi}\cos (At)\varphi_0+A^{1+\xi}\sin (At)\varphi_1\right\}ds\right\|_{L_{2, \gamma} (\mathbb{R}_+ ,H)}\hspace{3cm}\\
&\leq\left\|R_vA^{2+\xi}\left\{\gamma_k\{\cos (At)-\exp (-\gamma_kt)\}+A\sin (At)\right\}\varphi_0\right\|_{L_{2, \gamma} (\mathbb{R}_+ ,H)}+\\
&\hspace{1cm}+\left\|R_vA^{1+\xi}\left\{A\{\exp (-\gamma_kt)-\cos (At)\}+\gamma_k\sin (At)\right\}\varphi_1\right\|_{L_{2, \gamma} (\mathbb{R}_+ ,H)}\\
&= \left\|\cos (At) R_v\gamma_k A^{2+\xi}\varphi_0 \right\|_{L_{2, \gamma} (\mathbb{R}_+ ,H)}+  \left\|\mathrm{e}^{-\gamma_kt} R_v\gamma_k A^{2+\xi}\varphi_0 \right\|_{L_{2, \gamma} (\mathbb{R}_+ ,H)}+\\
&\hspace{1cm}+\left\|\sin (At)R_v A^{3+\xi}\varphi_0 \right\|_{L_{2, \gamma} (\mathbb{R}_+ ,H)}+\left\|\mathrm{e}^{-\gamma_kt} R_v A^{2+\xi}\varphi_1 \right\|_{L_{2, \gamma} (\mathbb{R}_+ ,H)}+\\
&\hspace{1cm}+\left\|\cos (At)R_v A^{2+\xi}\varphi_1 \right\|_{L_{2, \gamma} (\mathbb{R}_+ ,H)}+\left\|\sin (At)R_v\gamma_k A^{1+\xi}\varphi_1 \right\|_{L_{2, \gamma} (\mathbb{R}_+ ,H)}\\
&\lesssim \left\|\sum_{k=1}^{\infty}c_k\gamma_k \gamma^{-1}_k A^{-1}A^{2+\xi}\varphi_0 \right\|_{H}+\left\|\sum_{k=1}^{\infty}c_k\left(A^2+\gamma^2_k\right)^{-1} A^{3+\xi}\varphi_0 \right\|_{H}\\
&\hspace{1cm}+\left\|\sum_{k=1}^{\infty}c_k \left(A^2+\gamma^2_k\right)^{-1}A^{2+\xi}\varphi_1\right\|_{H}+\left\|\sum_{k=1}^{\infty}c_k\gamma_k \gamma^{-1}_k A^{-1}A^{1+\xi}\varphi_1 \right\|_{H}\\
&\lesssim 2\sum_{k=1}^{\infty}c_k \left(\|A^{1+\xi}\varphi_0\|_{H}+\|A^{\xi}\varphi_1\|_{H}\right)
\end{align*}} where $R_v=\sum_{k=1}^{\infty}c_k\left(A^2+\gamma^2_k\right)^{-1}$. Thus, from (\ref{desired-inequality}) and the last estimate, we get
{\small \begin{align} \label{estimate-omega}
\|\omega\|_{W^2_{2, \gamma} (\mathbb{R}_+ ,H)}&\leq d\|A^{2-\xi}f_1\|_{L_{2, \gamma} (\mathbb{R}_+ ,H)}\leq d\left(\|A^{2-\xi}f(t)\|_{L_{2, \gamma} (\mathbb{R}_+ ,H)}+\|A^{1+\xi}\varphi_0\|_{H}+\|A^{\xi}\varphi_1\|_{H}\right).
\end{align}} The estimation of $\|v(t)\|_{W_{2, \gamma}^{n}(\mathbb{R}_{+}, A^2)}:=\|\cos(At)\varphi_0\|_{W_{2, \gamma}^{n}(\mathbb{R}_{+}, A^2)}+\|A^{-1}\sin (At)\varphi_1\|_{W_{2, \gamma}^{n}(\mathbb{R}_{+}, A^2)}$ is obtained inmediatly. Indeed,   
{\small \begin{align*}
\|\cos(At)\varphi_0\|_{W_{2, \gamma}^{n}(\mathbb{R}_{+}, A^2)} &= \left(\int_{0}^{\infty}e^{-2\gamma t}\left(\|A^2\cos(At)\varphi_0\|^{2}_{H}+\|A^2 \cos(At)\varphi_0\|^{2}_{H}\right) dt\right)^{1/2}\nonumber\\
&<\left(2\int_{0}^{\infty}e^{-2\gamma t}\left(\|A^2\varphi_0\|^{2}_{H}\right) dt\right)^{1/2}<\frac{1}{\gamma^{1/2}}\|A^2\varphi_0\|_{H}\\
\|A^{-1}\sin(At)\varphi_1\|_{W_{2, \gamma}^{n}(\mathbb{R}_{+}, A^2)} &= \left(\int_{0}^{\infty}e^{-2\gamma t}\left(\|A\sin(At)\varphi_1\|^{2}_{H}+\|A \sin(At)\varphi_1\|^{2}_{H}\right) dt\right)^{1/2}\nonumber\\
&<\left(2\int_{0}^{\infty}e^{-2\gamma t}\left(\|A\varphi_1\|^{2}_{H}\right) dt\right)^{1/2}<\frac{1}{\gamma^{1/2}}\|A\varphi_1\|_{H} 
\end{align*}} This mean that $\varphi_0 \in \dom (A^2)=H_2$ and $\varphi_1 \in \dom(A)=H_1$. Hence, 
from (\ref{estimate-omega}) and obtained above it follows that 
\begin{align*} 
\|u(t)\|_{W^2_{2, \gamma} (\mathbb{R}_+ ,H)}\leq d\left(\|A^{2-\xi}f(t)\|_{L_{2, \gamma} (\mathbb{R}_+ ,H)}+\|A^{2}\varphi_0\|_{H}+\|A\varphi_1\|_{H}\right)\hspace{0.2cm} \mbox{for all $\xi \in [0, 1]$}. 
\end{align*}  with a constant $d$ that does not depend on the vector-valued function $f$ and the vectors $\varphi_0$, $\varphi_1$. Therefore, we have obtained the first estimate of Theorem \ref{theorem-about-solvability}.

2. Suppose, now, that the condition (\ref{condition-5}) does not hold. Then (\ref{condition-4}) and (\ref{expression-function-h}) imply that
\begin{align*}
\|A^{2-\xi}h(t)\|_{L_{2, \gamma} (\mathbb{R}_+ ,H)}&= \left\|\int_{0}^{t}\sum_{k=1}^{\infty}c_k \mathrm{e}^{-\gamma_k(t-s)}\left\{A^{2+\xi}\cos (At)\varphi_0+A^{1+\xi}\sin (At)\varphi_1\right \}ds \right\|_{L_{2, \gamma} (\mathbb{R}_+ ,H)}\\
&\leq \frac{1}{\sqrt{2\gamma}}\sum_{k=1}^{\infty}\frac{c_k}{\gamma_k}\left(\|A^{2+\xi}\varphi_0\|_{H}+\|A^{1+\xi}\varphi_1\|_{H}\right). 
\end{align*} Therefore, for all $\xi \in [0, 1]$ we have
\begin{align*}
\|\omega\|_{W^2_{2, \gamma} (\mathbb{R}_+ ,H)}\leq d\|A^{2-\xi}f_1\|_{L_{2, \gamma} (\mathbb{R}_+ ,H)}\leq d\left(\|A^{2-\xi}f(t)\|_{L_{2, \gamma} (\mathbb{R}_+ ,H)}+\|A^{2+\xi}\varphi_0\|_{H}+\|A^{1+\xi}\varphi_1\|_{H}\right). 
\end{align*} Consequently, the second estimate of Theorem \ref{theorem-about-solvability} is calculated:
\begin{align*}
\|u(t)\|_{W^2_{2, \gamma} (\mathbb{R}_+ ,H)}\leq d\left(\|A^{2-\xi}f(t)\|_{L_{2, \gamma} (\mathbb{R}_+ ,H)}+\|A^{2+\xi}\varphi_0\|_{H}+\|A^{1+\xi}\varphi_1\|_{H}\right) \hspace{0.2cm}\mbox{for all $\xi \in (0, 1]$},
\end{align*}  with a constant $d$ that does not depend on the vector-valued function $f$ and the vectors $\varphi_0$, $\varphi_1$. 
\section{Comments and observations}
Let us mention some results closely related to those obtained in the present paper. In \cite{{VW}, {KVW}} well-defined solvability is studied for problems of the form (\ref{solvability-solution-1}) and (\ref{solvability-solution-2}) under the hypothesis that the kernel $K(t)$ is of class $W_{1}^{1}(\mathbb{R}_+)$. In \cite{VR} and in the presente paper was included the case when $K(t) \notin W_{1}^{1}(\mathbb{R}_+)$ (see the second estimate of Theorem \ref{theorem-about-solvability}).  

The proof in \cite{{VW}, {KVW}}  about the existence theorem essentially utilizes the decay to the Laplace transform $\widehat{K}(\zeta)$, but does not involve the assumption that the function $K(t)$ can be represented as an exponential series of the form (\ref{suma-series-exponential}) (case that was considered in \cite{VR} and in the present work). It is noteworthy here that the correct solvability of hyperbolic Volterra equations considered here is a general and natural extension of what was presented in \cite{VR}. 

Hereinafter, we provide some corrections and accuracies of paper \cite{RPV}. 

\begin{lemma}\label{lemma-de-ceros-finitos} Consider the function 
\begin{align}
\ell_{n,N}(\zeta)= {\zeta}^2+ a^{2}_{n} \left(1-\frac{1}{a^{2(1-\xi)}_{n}}\sum_{k=1}^{N}\frac{c_k}{\zeta+\gamma_k}\right):= {\zeta}^2+a^{2}_{n}f_{n, N}(\zeta),\hspace{1cm} k=1,...N.
\end{align}
 Then the zeroes of the function $\ell_{n,N}$ form a set of real zeroes  $\{\mu_{n, k}\}^{N}_{k=1}$, such that
{\small \begin{align}
   -\gamma_k<\mu_{n, k} &<x_{n, k} < -\gamma_{k-1}<\mu_{n, k-1}<x_{n, k-1}<\cdots < -\gamma_{1}<\mu_{n, 1} <x_{n, 1}<0,\label{desigualdad-ceros}\\
  \mu_{n, k}-x_{n, k}&=O\left(\frac{1}{a^{2(1-\xi)}_{n}}\right), \hspace{0.2cm}0\leq\xi<1/2,\hspace{0.2cm} k=1,2,\cdots N,\hspace{0.2cm} a_{n} \to \infty,\nonumber\\
\mu_{n, k}-x_{n, k}&=O\left(\frac{1}{a_{n}}\right), \hspace{1cm}\xi=1/2,\hspace{0.9cm} k=1,2,\cdots N,\hspace{0.2cm} a_{n} \to \infty,\nonumber\\
\mu_{n, k}-x_{n, k}&=O\left(\frac{1}{a^{2\xi}_{n}}\right), \hspace{0.9cm}1/2<\xi\leq1,\hspace{0.2cm} k=1,2,\cdots N,\hspace{0.2cm} a_{n} \to \infty,\nonumber
  \end{align}} where $x_{n, k}$ are real zeroes of the function $f_{n, N}(\zeta)$ together with a pair of complex-conjugate zeroes $\mu^{\pm}_{n}$,  $\mu^{+}_{n}= \overline{\mu^{-}_{n}}$, which admit the asymptotic representation for all $\xi \in (0, 1)$ and $a_{n} \to \infty$
 {\small \begin{align} 
&\mu^{\pm}_{n}(\xi)=-\frac{1}{2} \frac{1}{a_{n}^{2(1-\xi)}}\sum_{j=1}^{N}c_j+O\left(\frac{1}{a^{2(1-\xi)}_{n}}\right)\pm i\left(a_{n}+O\left(\frac{1}{a^{1-2\xi}_{n}}\right)\right),  \hspace{0.2cm}&\mbox{$0\leq \xi<\frac{1}{2}$},\label{asimptota-finita1}\\
&\mu^{\pm}_{n}(\xi)=-\frac{1}{2} \frac{1}{a_{n}}\sum_{j=1}^{N}c_j+O\left(\frac{1}{a_{n}}\right)\pm i\left(a_{n}+O\left(1\right)\right),  \hspace{0.2cm}&\mbox{$\xi=\frac{1}{2}$},\label{asimptota-finita2}\\
&\mu^{\pm}_{n}(\xi)=-\frac{1}{2} \frac{1}{a_{n}^{2(1-\xi)}}\sum_{j=1}^{N}c_j+O\left(\frac{1}{a^{2\xi}_{n}}\right)\pm i\left(a_{n}+O\left(\frac{1}{a^{2\xi-1}_{n}}\right)\right),  \hspace{0.2cm}&\mbox{$\frac{1}{2}<\xi\leq 1$}\label{asimptota-finita3} 
\end{align}} 
\end{lemma}

In \cite{RPV} was assumed the following condition holds: 
\begin{align} \label{suprem-de-las-gammas}
\sup_{k}\{\gamma_k(\gamma_{k+1}-\gamma_k)\}=+\infty
\end{align}

\begin{theorem}\label{teorema-de-ceros-numerables} Suppose that conditions (\ref{condition-5}) and (\ref{suprem-de-las-gammas}) are satisfied. Then the zeros of the meromorphic function  $\ell_{n}(\zeta)$ form a countable set of real zeroes  $\{\mu_{n, k}\}^{\infty}_{k=1}$, which satisfy
\begin{align}
   &-\gamma_k<\mu_{n, k} < -\gamma_{k-1}, \hspace{0.4cm}\lim_{n\to\infty} \mu_{n, k}=-\gamma_k,\label{inequality-sum-finit}   
  \end{align}
together with a pair of complex-conjugate zeroes $\mu^{\pm}_{n}$,  $\mu^{+}_{n}= \overline{\mu^{-}_{n}}$ which admit the following asymptotic representation
  {\small \begin{align} 
&\mu^{\pm}_{n}(\xi)=-\frac{1}{2} \frac{1}{a_{n}^{2(1-\xi)}}\sum_{j=1}^{\infty}c_j+O\left(\frac{1}{a^{2(1-\xi)}_{n}}\right)\pm i\left(a_{n}+O\left(\frac{1}{a^{1-2\xi}_{n}}\right)\right),  \hspace{0.2cm}&\mbox{$0\leq \xi<\frac{1}{2}$}\label{asymptota-sum-finita-para-mu1}\\
&\mu^{\pm}_{n}(\xi)=-\frac{1}{2} \frac{1}{a_{n}}\sum_{j=1}^{\infty}c_j+O\left(\frac{1}{a_{n}}\right)\pm i\left(a_{n}+O\left(1\right)\right),  \hspace{0.2cm}&\mbox{$\xi=\frac{1}{2}$},\label{asymptota-sum-finita-para-mu2}\\
&\mu^{\pm}_{n}(\xi)=-\frac{1}{2} \frac{1}{a_{n}^{2(1-\xi)}}\sum_{j=1}^{\infty}c_j+O\left(\frac{1}{a^{2\xi}_{n}}\right)\pm i\left(a_{n}+O\left(\frac{1}{a^{2\xi-1}_{n}}\right)\right),  \hspace{0.2cm}&\mbox{$\frac{1}{2}<\xi\leq 1$} \label{asymptota-sum-finita-para-mu3}
\end{align}} 
\end{theorem} 
Under the conditions of Theorem \ref{teorema-de-ceros-numerables}, the spectrum $\sigma(L)$ of the operator-valued function $L(\zeta)$ coincides with the set of zeroes $\{\mu_{n, k}\}^{\infty, \infty}_{n, k=1}$ and  $\{\mu^{\pm}_{n}\}^{\infty}_{n=1}$ of function $\ell_n(\zeta)$, that is, 
\begin{align*}
 \sigma(L)=\overline{\left(\cup^{\infty}_{n,k=1}\mu_{n, k}\right) \cup \left(\cup^{\infty}_{n} \mu^{\pm}_{n}\right)}
\end{align*} 

\begin{condition} \label{condition-1}In the case when the condition  (\ref{condition-5}) is not satisfied, the sequences $\{c_{k}\}^{\infty}_{k=1}$, $\{\gamma_{k}\}^{\infty}_{k=1}$ have the following asymptotic representation 
   $c_k=\frac{A}{k^{\alpha}} + O\left( \frac{1}{k^{\alpha +1}}\right)$, $\gamma_k=B k^{\beta} + O( k^{\beta-1})$, $(k \to \infty)$ where  $c_k>0$, $\gamma_{k+1}>\gamma_{k}>0$, constants $A>0, B>0$,  $0<\alpha \leq 1$, $\alpha +\beta>1$ such that $\sum_{k=1}^{\infty}\frac{c_k}{\gamma_k}<1$. 
 
If the condition  $\gamma_{k+1}-\gamma_{k} \thickapprox k^{\beta-p}$ is satisfied for some $p \in \mathbb{N} $, $p \geq 1$ and $k \to \infty$ then the condition  (\ref{suprem-de-las-gammas}) of Ivanov is satisfied when $p <2\beta$. 
\end{condition}

\begin{theorem} \label{teorema-de-ceros-infinitos} Assume that the Condition \ref{condition-1} is satisfied. Then the zeroes of the function $\ell_{n}(\zeta)$ form a countable set of real zeroes $\{\lambda_{n, k}|k \in \mathbb{N}\}$  such that 
 \begin{align} \label{desigualdad-ceros-infinitos}
  \cdots -\gamma_k<-\lambda_{n,k}<\cdots<-\gamma_1<0, \hspace{0.3cm}\lim_{n \to \infty}  \lambda_{n,k}=-\gamma_k
  \end{align} together with a pair of complex-conjugate zeroes $\lambda^{\pm}_{n}$,  $\lambda^{+}_{n}= \overline{\lambda^{-}_{n}}$ which admit the asymptotic representations when $a_n \to \infty$:
 {\small \begin{align}
  &\lambda^{\pm}_{n}(\xi)=\frac{AD_1B^{r-1}}{\beta a^{r+1-2\xi}_n}+O\left(\frac{1}{a^{\delta_1}_n}\right)\pm i\left(a_n+\frac{AD_2B^{r-1}}{\beta a^{r+1-2\xi}_n}+O\left(\frac{1}{a^{\delta_2}_n}\right)\right), &\mbox{$r\in \left(0, \frac{1}{2}\right)$}, \label{aimptotas-de-ceros-infinitos-1} \\
  &\lambda^{\pm}_{n}(\xi)=\frac{AD_1B^{r-1}}{\beta a^{r+1-2\xi}_n}+O\left(\frac{1}{a^{2(1-\xi)}_n}\right)\pm i\left(a_n+\frac{AD_2B^{r-1}}{\beta a^{r+1-2\xi}_n}+O\left(\frac{1}{a^{\delta_2}_n}\right)\right), &\mbox{$r \in \left[\frac{1}{2}, 1\right)$}, \label{aimptotas-de-ceros-infinitos-2} \\
  &\lambda^{\pm}_{n}(\xi)=-\frac{1}{2}\frac{A}{\beta}\frac{\ln a_n}{a^{2(1-\xi)}_n}+O\left(\frac{1}{a^{2(1-\xi)}_n}\right)\pm ia_n,&\mbox{$r=1$}, \label{aimptotas-de-ceros-infinitos-3} 
  \end{align}} where $\delta_1= \min \{2(1-\xi), 2r+3-4\xi\}$, $\delta_2=2r+3-4\xi$ $k=1, 2$, $r:=\frac{\alpha+\beta-1}{\beta}$, the constant $D=D_1+iD_2$ depends of $r$ and  is defined as follows: 
 {\small
 \begin{align*} 
D:=\frac{i}{2}\int_{0}^{\infty}\frac{dt}{t^r(i+t)}=\frac{1}{2}\left(\int_{0}^{\infty}\frac{dt}{t^{r}(1+t^{2})}+i\int_{0}^{\infty}\frac{dt}{t^{r-1}(1+t^{2})}\right)=\frac{\pi}{2}\frac{\exp{\left(i\frac{\pi}{2}(1-r)\right)}}{\sin(\pi r)}. 
  \end{align*}}
  \end{theorem}
Under the conditions of Theorem \ref{teorema-de-ceros-infinitos} , the spectrum $\sigma(L)$ of the operator-valued function $L(\zeta)$ coincides with the set of zeroes $\{\lambda_{n, k}\}^{\infty, \infty}_{n, k=1}$ and  $\{\lambda^{\pm}_{n}\}^{\infty}_{n=1}$ of function $\ell_n(\zeta)$, that is, 
\begin{align*}
 \sigma(L)=\overline{\left(\cup^{\infty}_{n,k=1}\lambda_{n, k}\right) \cup \left(\cup^{\infty}_{n} \lambda^{\pm}_{n}\right)}
\end{align*}

\end{document}